\documentstyle[12pt,epsf,rotate,a4]{article}

\def\exp{{\rm e}}
\def\log{{\rm log}}
\def\Tr{{\rm Tr}}

\begin{document}

\baselineskip=24pt

\title{
\vspace{-3.0cm}
\begin{flushright}
{\normalsize August, 1995}\\
\vspace{-0.3cm}
{\normalsize UTHEP-312}\\
\vspace{-0.4cm}
{\normalsize hep-th/9508026}\\
\end{flushright}
\vspace*{2.0cm}
{\Large {\bf Non Scale-Invariant\\  Topological Landau-Ginzburg Models}}
\bigskip
\bigskip
\bigskip
\bigskip
       }

\author{ Masayuki Noguchi and Sung-Kil Yang \\ \bigskip \\
        \it{Institute of Physics,
            University of Tsukuba, Ibaraki 305, Japan}}
\date{}
\maketitle

\vspace*{0.5cm}

\begin{abstract}

The Landau-Ginzburg formulation of two-dimensional topological sigma models
on the target space with positive first Chern class is considered. The
effective Landau-Ginzburg superpotential takes the form of logarithmic type
which is characteristic of supersymmetric theories with the mass gap.
The equations of motion yield the defining relations of the quantum cohomology
ring. Topological correlation functions in the $CP^{n-1}$ and Grassmannian
models are explicitly evaluated with the use of the logarithmic
superpotential.

\end{abstract}

\newpage
%
%

It is well-known that the target space of $N=2$ supersymmetric sigma models
in two dimensions is a K\"ahler manifold \cite{Zumino}.
The coefficient of the one-loop
$\beta$-function is then fixed by the first Chern class $c_1$ of K\"ahler
manifolds.
For $c_1 >0$, which is the case of the $CP^{n-1}$ and the Grassmannian, the
$N=2$ sigma models are asymptotically free and possess the dynamically
generated mass gap. As emphasized in \cite{Witten4}
the $N=2$ models with the mass gap
reduce at large distances to topological field theories which capture a
non-trivial dynamics of the supersymmetric vacua. An important manifestation
of such property is the quantum cohomology ring of K\"ahler manifolds
\cite{Witten2,V}. Another characteristic aspect of supersymmetric theories
with the mass gap is observed in the effective superpotential. It is seen
that the {\it{logarithmic}} superpotential
is ubiquitous in the low-energy effective description of supersymmetric,
asymptotically free theories \cite{VY}.

In this paper we study the two-dimensional topological sigma model
with $c_1>0$, which is obtained by twisting the corresponding $N=2$ sigma
model \cite{Witten2,Witten1}.
Our point is to employ the logarithmic effective superpotential
upon evaluating correlation functions. We first discuss the  $CP^{n-1}$
model for which the logarithmic effective superpotential has been known
for a long time  \cite{DDDS}. In a more conventional approach, on the
other hand, the same class of topological correlation functions
has already been calculated in the
Landau-Ginzburg (LG) formulation of perturbed
$N=2$ superconformal theories \cite{I}.
These theories are completely characterized by  polynomial
superpotentials. Comparing the results obtained in both ways we shall discuss
the relation between the two approaches.
Then we extend our considerations to the Grassmannian models,
and will clarify how
logarithmic superpotentials serve as effective superpotentials in
topological field theory.

%
%

We start with the $N=2$ sigma model whose target space is the
complex $(n-1)$-dimensional projective space $CP^{n-1}$.
The Lagrangian of the $CP^{n-1}$ model is given by
\begin{equation}
L=\int d^4 \theta \left[
\sum_{i=1}^n{\bar{\Phi}_i} {\exp}^{-V} \Phi_i + {n\over{2g}} V
                  \right],
\end{equation}
in two-dimensional $N=2$ superspace,
where $\Phi_i$ are chiral superfields
and $g$ is a dimensionless coupling constant.
$V$ is an auxiliary $N=2$ vector superfield by virtue of which we
see manifest $U(1)$ gauge symmetry which is otherwise hidden.

If we eliminate the auxiliary field $V$ by
means of its equation of motion we have the standard form of $N=2$
sigma model.
In an interesting paper \cite{DDDS} (see also \cite{CV}), however,
the matter fields $\Phi_i$ are integrated out. Then one ends up with
the effective action
\begin{equation}
S_{eff} = {1 \over 4\pi}\int d^2 x \left[
                           \int d^2 \theta W_{log}(\lambda) + h.c. + \cdots
                     \right],
\label{cpea}
\end{equation}
where $\lambda$ is a gauge invariant, twisted chiral superfield
expressed in terms of the superderivative $\lambda = D_L {\bar D}_R V$
and the ellipses stand for possible $D$-term contributions
whose detail is irrelevant to our consideration in topological
field theory. The explicit form of the superpotential
reads \cite{DDDS,CV}
\begin{equation}
W_{log}(\lambda) =   \lambda
         \left( {\log} {\lambda^{n} \over{\mu^{n}}} - n \right) ,
\label{cpwlog}
\end{equation}
where $c_1=n$ and $\mu$ is a dynamically generated mass scale.

The logarithmic superpotential (\ref{cpwlog}) has the
following two important properties:
First the effective action (\ref{cpea}) produces
the correct anomaly structure of the $N=2$ $CP^{n-1}$ model.
An axial, conformal and $\gamma$ trace anomaly are all
reproduced by taking appropriate variations of $S_{eff}$ \cite{DDDS}.
Second the equation of motion $\partial_{\lambda} W_{log}(\lambda) = 0$
for $\lambda$ gives rise to
\begin{equation}
\lambda^n = \mu^n.
\label{lambda}
\end{equation}
This is nothing but the relation for the quantum cohomology ring of
$CP^{n-1}$ under the identification of $\lambda$ with the
harmonic one-form of $CP^{n-1}$ \cite{Witten2,V}.
Thus we observe that the potential (\ref{cpwlog}) is equipped with
the desired property required to be the effective potential for
the topological $CP^{n-1}$ model.

Before turning to the calculation of topological correlation functions
with the use of (\ref{cpwlog}) we review how a suitably
perturbed superpotential in the LG formulation of $N=2$
superconformal field theory works in describing the
$CP^{n-1}$ quantum cohomology ring. The $A_n$-type LG superpotential
is a polynomial of a single chiral superfield $X$ and takes the form
\cite{M,VW}
\begin{equation}
W(X) = {X^{n+1}\over{n+1}} .
\label{upcppol}
\end{equation}
The equation of motion
for $X$ obtained from (\ref{upcppol}) may be interpreted as the
relation for the {\it classical} cohomology ring of $CP^{n-1}$
if we again identify $X$ with the harmonic one-form
of $CP^{n-1}$. To generate the {\it{quantum}} cohomology ring of $CP^{n-1}$
the potential (\ref{upcppol}) needs to be perturbed by the most
relevant operator \cite{I}. We have
\begin{equation}
W_{pol}(X) = {X^{n+1}\over{n+1}} - {\beta} X ,
\label{cpwpol}
\end{equation}
where $\beta$ is a perturbation parameter.
The equation of motion $\partial_X W_{pol}(X)=0$ then yields the relation
\begin{equation}
X^n=\beta.
\end{equation}
Hence, as long as the cohomology ring structure is
concerned, both potentials $W_{log}$ in (\ref{cpwlog})
and $W_{pol}$ in (\ref{cpwpol}) work well and we see the correspondence
$\lambda \leftrightarrow X$, $\mu^n \leftrightarrow \beta.$

Let us now evaluate topological correlation
functions defined on the genus $g$ Riemann surface
 using the residue formula \cite{DVV,Vafa}
\begin{equation}
\langle F \rangle = \sum_{dW=0} F H^{g-1},
\label{corfunc}
\end{equation}
where $F$ is a function of topological
observables and $H$ is the Hessian
of superpotential.
Here the sum is taken over all the critical points of $W$ where $dW=0$.
It is trivial that the critical points are located at
$\lambda=\mu \exp^{{2\pi i\over n}j}$
$(X=\beta^{1/n}  \exp^{{2\pi i\over n}j})$
for $W_{log}$ $(W_{pol})$
with $j=0,1,\cdots,n-1$.
Thus both potentials have $n$-fold degenerate vacua. This is the
correct degeneracy since the Witten index of $CP^{n-1}$ is
equal to the Euler characteristic $\chi$ which is known to be
$\chi =n$ \cite{Witten3}. Substituting the  Hessians given by
$\partial_{\lambda}^2 W_{log} = n/\lambda$ and
$\partial_X^2 W_{pol} = n X^{n-1}$ it is straightforward
to obtain
\begin{equation}
\langle \lambda^m \rangle_{log}
= \sum_{dW_{log}=0} \lambda^m (n /\lambda)^{g-1}
=n^g \mu^{n(k+1-g)},
\label{corlog}
\end{equation}
and
\begin{equation}
\langle X^m \rangle_{pol} = \sum_{dW_{pol}=0} X^m (n X^{n-1})^{g-1}
=n^g \beta^{k},
\label{corpol}
\end{equation}
where $m = nk + (n - 1)(1-g)$.
Notice that $\langle \lambda^m \rangle =
\langle X^m \rangle =0$ if $m \not= nk + (n - 1)(1-g)$.

In both results correlation functions are
nonvanishing only when
$m=nk+(n-1)(1-g)$ where $k$ is understood as the degree of holomorphic
maps from the genus $g$ Riemann surface to $CP^{n-1}$.
This is the well-known $U(1)$ charge conservation law
in the topological sigma model
(without coupling to topological gravity);
$\sum_\alpha q_\alpha =d(1-g)+kc_1$ where $q_\alpha$ are $U(1)$ charges and
 $d$ is the complex dimension of
target manifold \cite{Witten2,Witten1}.
Under the identification $\beta = \mu^n$, however,
we observe that (\ref{corlog}) and (\ref{corpol})
differ by a factor of $\mu^{n(1-g)}$.
Otherwise the logarithmic potential yields the right result
derived by using $W_{pol}$.
In oder to understand this discrepancy properly
we have to remember that the Hessian is identified as the top
element $\phi_{top}$ of the chiral ring \cite{W}.
Since $\phi_{top}$ is the operator corresponding to the spectral flow,
the insertion of $H$ in (\ref{corfunc}) is responsible for realizing
the correct $U(1)$ charge conservation \cite{Vafa}.
For $W_{pol}$ we have $H_{pol}=nX^{n-1} = \phi_{top}$, whereas for
$W_{log}$ we have $H_{log}=n/\lambda$ which is not quite the form
of $\phi_{top}=\lambda^{n-1}$.
Noting the relation (\ref{lambda}), however, we find
$H_{log}\simeq \mu^{-n} \lambda^{n-1} = \mu^{-n} \phi_{top}$. This is a
simple, but key observation in the present paper, based on which we
understand why (\ref{corlog}) and (\ref{corpol}) differ by a factor
of $\mu^{n(1-g)}$. Notice also that $n$ is the first
Chern class of $CP^{n-1}$ which determines the one-loop $\beta$-function,
and hence controls the size of scaling violation.

%
%

Let us now turn to the $N=2$ Grassmannian model.
The Grassmannian $Gr(N,M)$ is defined to be the
set of complex $N$-dimensional linear subspaces of
a $(N+M)$-dimensional complex vector space, and as a homogeneous space,
$Gr(N,M)$ is expressed as $U(N+M)/U(N)\times U(M)$,
which is a natural extension of $CP^{n-1}$. Actually $Gr(1,n-1)$ is
nothing but $CP^{n-1}$.
In $N=2$ superspace formalism, the Lagrangian of the Grassmannian model
takes the form
\begin{equation}
L= \int d^4 \theta \left[
   \sum_{i=1}^{N+M} \sum_{a,b=1}^N
   {\bar{\Phi}}_i^a (\exp^{-V})^{ab} \Phi_i^b
   + {N+M \over 2g} \Tr \, V
                   \right] ,
\label{laggra}
\end{equation}
where
$\Phi$ is a matrix chiral superfield and
$V$ is an $N\times N$ matrix-valued $U(N)$ vector superfield.

In contrast to the $CP^{n-1}$ model, the explicit calculation of an
effective superpotential for the Grassmannian is a formidable task,
though (\ref{laggra}) is quadratic with respect to $\Phi$.
We have to look for possible indirect ways to find
the effective potential. One way is to examine
to what extent one can control the form of the superpotential
under the requirement of $N=2$ supersymmetry, $U(N)$ gauge symmetry
and the correct anomaly structure.
We were able to write down a few candidate superpotentials
which are $N=2$ supersymmetric as well as $U(N)$ gauge invariant
and yield desired anomalies. However these trial potentials
do not have the correct Witten index, i.e.
the Euler characteristic for $Gr(N,M)$.

To overcome this difficulty we follow another
route as discussed at length in \cite{CV}.
A fundamental field variable to describe the
effective action is the field-strength superfield $\Lambda$
which is gauge covariant, rather than gauge invariant, in the
non-abelian case. The cohomology ring, on the other hand,
should be generated by the gauge invariant objects $X_i$
$(i=0,1,\cdots,N)$ with $X_0=1$. The relation between $\Lambda$ and $X_i$
is given by \cite{CV}
\begin{equation}
\det (1+t \Lambda) = 1 + \sum_{i=1}^N X_i t^i.
\label{cov}
\end{equation}
Furthermore, under the assumption that
$\Lambda$ and $\bar \Lambda$ commute,
$\Lambda$ may be reduced to the diagonal matrix,
$\Lambda= {\rm diag}(\lambda_1,\lambda_2,\cdots,\lambda_N)$.
This is the idea of abelianization in \cite{Witten4}.
Then the effective superpotential is written as \cite{CV,Witten4}
\begin{equation}
W_{log}=\sum_{a=1}^N \lambda_a
        \left(
              \log{\lambda_a^{N+M} \over \mu^{N+M}} -N -M
        \right),
\label{grwlog}
\end{equation}
where $N+M$ is the first Chern class of $Gr(N,M)$. In view of (\ref{cpwlog})
the effective potential (\ref{grwlog}) looks like $N$ copies of the
$CP^{N+M-1}$ model. The equation of motion $\partial_{\lambda_a}W_{log}=0$
gives us
\begin{equation}
\lambda_a^{N+M}=\mu^{N+M}, \hskip10mm a=1,2,\cdots ,N.
\label{grring}
\end{equation}
The quantum cohomology ring of $Gr(N,M)$ is generated by $\{X_i\}$
which are symmetric functions of $\lambda_a$ subject to (\ref{grring}).

Let us check the Witten index using (\ref{grwlog}).
When $\Lambda$ is reduced to the diagonal matrix the path integral measure
for $\Lambda$ becomes $[\prod_ad\lambda_a] \triangle (\lambda)^2$ where
the Vandermonde determinant
$\triangle (\lambda)=\prod_{a<b}(\lambda_a-\lambda_b)$ is the Jacobian
arising from the angular part. Thus we see that $\lambda_a$'s repel each
other, and hence in the vacuum
configuration $\{ \lambda_a\}$ we have $\lambda_a \not=\lambda_b$ where
$\lambda_a=\mu e^{2\pi j_a \over N+M}$ ($j_a=0,1,\cdots ,N+M-1$) from
(\ref{grring}).
Moreover $X_i$ is expressed as a symmetric
polynomial in $\lambda_a$, so we have the
permutation symmetry of $N$ objects.
Therefore the number of degenerate vacua turns out to be
\begin{equation}
{(M+N)(M+N-1)\cdots (M+1) \over N!} =
\left(
\begin{array}{c}
N+M\cr
N
\end{array}
\right).
\end{equation}
This is the Euler characteristic for $Gr(N,M)$,
so we find that (\ref{grwlog}) possesses the desirable
vacuum structure.

We briefly recall here the cohomology ring of $Gr(N,M)$
which is generated by $X_0,X_1,$ \\ $\cdots ,X_N$ where $X_j$ is
a $(j,j)$ form \cite{grcr}. First introduce the following polynomials
\begin{equation}
X^{(N)}(t) = 1 + \sum_{i=1}^N X_i t^i, \hskip10mm
Y^{(N)}(t) = (X^{(N)}(-t))^{-1} = \sum_{n\geq 0} Y^{(N)}_n t^n.
\end{equation}
Then the ideal of the classical $Gr(N,M)$ ring is given by
\begin{equation}
Y^{(N)}_n = 0, \hskip10mm  n = M+1,\cdots,M+N.
\label{clrel}
\end{equation}
Furthermore, let \cite{G}
\begin{equation}
W^{(N)}(t) = - \log(X^{(N)}(-t)) = \sum_{i\geq 0} W^{(N)}_i t^i \ ,
\end{equation}
then we have
\begin{equation}
{\partial W^{(N)}(t)\over {\partial X_i}} = - Y^{(N)}(t)(-t)^i,
\hskip10mm  i=1,2,\cdots,N,
\end{equation}
thereby
\begin{equation}
{\partial W_{N+M+1}^{(N)} \over \partial X_i}
=(-1)^{i+1}Y^{(N)}_{N+M+1-i}, \hskip10mm i=1,2,\cdots,N.
\end{equation}
Hence the generating function for the
classical relation (\ref{clrel}) is given by $W^{(N)}_{N+M+1}(X_i)$
which is also regarded as the (unperturbed)
LG superpotential for $Gr(N,M)$  $N=2$ superconformal theory \cite{W,G}.

To consider the quantum ring of $Gr(N,M)$ the superpotential
$W^{(N)}_{N+M+1}$ is
perturbed by the most relevant operator
$X_1$ \cite{I}. The deformed potential reads
\begin{equation}
W_{pol} \equiv
W_{N+M+1}(X_i) - \beta X_1 .
\label{grwpol}
\end{equation}
The equation of motion $\partial_{X_i} W_{pol} = 0$
gives
\begin{equation}
(-1)^{i+1}Y^{(N)}_{N+M+1-i}=\beta \delta_{i,1},
\label{Yring}
\end{equation}
which is the defining relation of the $Gr(N,M)$ quantum cohomology ring.

Let us write \cite{G,CV1,I}
\begin{equation}
X^{(N)}(t)=\prod_{a=1}^N(1+tq_a),
\end{equation}
then (\ref{grwpol}) is rewritten as
\begin{equation}
W_{pol}=\sum_{a=1}^N
\left(  {q_a^{N+M+1}\over{N+M+1}} - \beta q_a  \right).
\label{grwpolq}
\end{equation}
Using the chain rule $\partial_{X_i} W_{pol} = 0=
(\partial_{X_i}q_a) \partial_{q_a} W_{pol}$ it is seen that (\ref{Yring})
is obtained from
\begin{equation}
q_a^{N+M} = \beta ,\hskip10mm  a = 1,2,\cdots,N,
\label{grrel}
\end{equation}
where $q_a\not= q_b$ has been assumed so that
$\det (X_i/q_a)=\prod_{a<b}(q_a-q_b)$ is nonvanishing. Note here that
(\ref{grwpolq}) takes the form of $N$ copies of the deformed $A_{N+M}$-type
LG theory. Moreover, from (\ref{grring}) and (\ref{grrel}) we see the
important correspondence $\lambda_a \leftrightarrow q_a$,
$\mu^{N+M} \leftrightarrow \beta$.

Having two types of the potential (\ref{grwlog}) and (\ref{grwpol})
for the Grassmannian, we now wish to calculate topological correlation
functions with the aid of (\ref{corfunc}). For $W_{pol}$ we substitute
$H_{pol}(X) = (-1)^{N(N-1)/2}\det(\partial_i\partial_jW_{pol})$ with
$\partial_i \equiv \partial/ \partial X_i$, whereas for $W_{log}$ we have to
take into account the Jacobian factor as pointed out before, and thus
the ``effective'' Hessian $H_{log}$ is given by
$H_{log}(\lambda)=(-1)^{N(N-1)/2}\triangle (\lambda)^{-2}
\det(\partial_a\partial_bW_{log})$ with
$\partial_a \equiv \partial/ \partial \lambda_a$.
Let us examine the following two simple examples:

{\underline {$Gr(2,2)$ model}}\\
For the logarithmic potential, we get
\begin{equation}
\langle X_1^n X_2^m \rangle_{log} =
       (-1)^k 2^{2k-m+1+g}\mu^{4\{k+2(1-g)\}}
\label{gr22corfun1}
\end{equation}
for $n+2m=4k+4(1-g)$ with $n\neq 0$, and
\begin{equation}
\langle X_2^m \rangle_{log}       =
       [2^{2g-1} + (-1)^k 2^{3g-1}]\mu^{4\{k+2(1-g)\} }
\end{equation}
for $2m=4k+4(1-g)$, where
$X_1 = \lambda_1 + \lambda_2$ and $X_2 = \lambda_1  \lambda_2$.
On the other hand, employing the polynomial potential
\begin{equation}
W_{pol} = {X_1^5\over 5} - X_1^3 X_2 + X_1 X_2^2 - \beta X_1,
\label{gr22wpol}
\end{equation}
we evaluate correlation functions as
\begin{equation}
\langle X_1^n X_2^m \rangle_{pol} =
       (-1)^k 2^{2k-m+1+g}\beta^{k}
\end{equation}
for $n+2m=4k+4(1-g)$ with $n\neq 0$, and
\begin{equation}
\langle X_2^m \rangle_{pol}       =
      [2^{2g-1} + (-1)^k 2^{3g-1}] \beta^{k}
\label{gr22corfun2}
\end{equation}
for $2m=4k+4(1-g)$.
In (\ref{gr22corfun1})-(\ref{gr22corfun2}) correlation
functions vanish if $n+2m \neq 4k + 4(1-g)$.
For both types of the potential we confirm the correct $U(1)$ charge
selection rule with degree $k$ instantons
of the topological Grassmannian model, though there exists a discrepancy by
a factor of $\mu^{8(1-g)}$
(putting $\beta =\mu^4$) as in the case of $CP^{n-1}$.
Let us next proceed to more interesting example.

{\underline{$Gr(2,3)$ model}}\\
After some algebra we obtain
correlation functions for the logarithmic potential
\begin{equation}
\langle X_1^n \rangle_{log} =
       5(5\sqrt{5})^{g-1}\mu^{5\{k+2(1-g)\} }
             \left[
                     \left( {\sqrt{5}-1\over2} \right)^{5(k-g+1)} +
              (-1)^k \left( {\sqrt{5}+1\over2} \right)^{5(k-g+1)}
             \right]
\label{grcorlog}
\end{equation}
for $n = 5k + 6(1-g)$,
where $X_1 = \lambda_1 + \lambda_2$. For the polynomial potential
\begin{equation}
W_{pol} =
{X_1^6 \over 6} - X_1^4 X_2 + {3\over 2} X_1^2 X_2^2 -
{X_2^3 \over 3} - \beta X_1,
\label{gr23wpol}
\end{equation}
we get
\begin{equation}
\langle X_1^n \rangle_{pol} =
       5(5\sqrt{5})^{g-1}\beta^{k}
             \left[
                     \left( {\sqrt{5}-1\over2} \right)^{5(k-g+1)} +
              (-1)^k \left( {\sqrt{5}+1\over2} \right)^{5(k-g+1)}
             \right]
\label{grcorpol}
\end{equation}
for $n = 5k + 6(1-g)$. Replacing $\beta$ by $-\beta$
reproduces the result derived earlier in \cite{I}.
In (\ref{grcorlog}) and (\ref{grcorpol}) $\langle X_1^n \rangle=0$
if $n \neq 5k+6(1-g)$.
Thus, the situation is similar to the $Gr(2,2)$ and $CP^{n-1}$
models, but this time
the difference is by a factor of $\mu^{10(1-g)}$ (putting $\beta =\mu^5$).

The origin of the discrepancy observed above is figured out by examining
the relation between the Hessian $H$ and the top element $\phi_{top}$ of
the chiral ring as in the previous case of $CP^{n-1}$. For $W_{log}$ we have
\begin{equation}
H_{log}(\lambda) = (-1)^{N(N-1)\over 2} \triangle(\lambda)^{-2}\det
\left(
           {\partial^2 W_{log}\over \partial \lambda_a\partial \lambda_b}
\right)
= {(-1)^{N(N-1)\over 2}
  (N+M)^N \over \prod_{a<b}(\lambda_a-\lambda_b)^2 \prod_{a=1}^N \lambda_a},
\end{equation}
while for $W_{pol}$ we get
\begin{equation}
H_{pol}(X)|_{dW=0} = (-1)^{N(N-1)\over 2}
\det \left(
           {\partial^2 W_{pol}\over \partial X_i\partial X_j}
     \right) {\Bigg|}_{dW=0}
= (-1)^{N(N-1)\over 2}{(N+M)^N \prod_{a=1}^N q_a^{N+M-1}
     \over \prod_{a<b}(q_a-q_b)^2}.
\end{equation}
Under the correspondence $\lambda_a \leftrightarrow q_a$ we find
\begin{equation}
{H_{pol}(X) \over H_{log}(\lambda)} {\Bigg|}_{dW=0}
\simeq \prod_{a=1}^N \lambda_a^{N+M}
=\mu^{N(N+M)},
\label{ratio}
\end{equation}
where we have used the relation (\ref{grring}) of the quantum ring.
Since $H_{pol}(X)=\phi_{top}$ in the polynomial-type LG description,
it follows from (\ref{ratio}) that
$H_{log}(\lambda)\simeq \mu^{-N(N+M)} \phi_{top}$. This explains why we
encountered the extra factor $\mu^{8(1-g)}$ ($\mu^{10(1-g)}$)
for the $Gr(2,2)$
($Gr(2,3)$) model with $W_{log}$. Thus, whenever we use
the topological residue formula in the theory with
logarithmic potential which is characteristic of scaling violation,
we learn that $H_{log}$ is replaced by $\mu^{rc_1}H_{log}$
where $r$ is the number of fundamental LG fields and $c_1$ is the first
Chern class.

We have presented a LG formulation of the topological sigma model.
The logarithmic effective superpotential we have used was obtained by
exact path-integral computations for sigma models \cite{DDDS,CV}.
Thus our LG model is exactly equivalent to the sigma model as stressed
in \cite{CV}. It will be interesting to develop the LG description of $N=2$
sigma models on other homogeneous spaces than Grassmannian.
In concluding this paper let us finally remark the following two points:

1)\, We have seen that the topological $CP^{n-1}$ model without coupling to
topological gravity is well formulated as LG models in terms of either
$W_{log}$ (\ref{cpwlog}) or $W_{pol}$ (\ref{cpwpol}). When coupling to
topological gravity, however, the situation changes drastically. After
coupling to gravity the LG model with the polynomial superpotential
(\ref{cpwpol}) describes the minimal model with the topological central
charge $\hat c <1$, rather than the $CP^{n-1}$ model with $\hat c=d=n-1$.
In fact, for the $CP^1$ model coupled to gravity, it was found in \cite{EHY}
that a suitable LG superpotential takes the form of exponential
interactions; $\mu(e^X+e^{-X})$ with $X$ being the LG field. Then it might
be asked if one could find any place where the logarithmic superpotential
plays a role. At present we cannot answer to this question, but it is worth
pointing out that the action of the $CP^1$ matrix model looks
analogous to $W_{log}$ \cite{EY,EHY}.
Precise reasoning for this similarity is desirable.

2)\, There exists another interesting class of non scale-invariant
topological field theories. Let us take the superpotential with
exponential interactions \cite{CV}
\begin{equation}
W_{exp}=\mu \left( {1 \over n-1}e^{(n-1)X}+e^{-X} \right) ,
\end{equation}
where $\mu$ is a mass scale and $n=2,3,\cdots$.
Let $\mu=e^{t/c_1}$ with $c_1=n/(n-1)$ and make a shift
of the LG field $X \rightarrow X-t/n$, then we have
\begin{equation}
W_{exp}= {1 \over n-1}p^{n-1}+e^{t}p^{-1},
\label{lax}
\end{equation}
where we have put $p=e^X$. The superpotential in this form can be regarded as
the Lax operator of a particular reduction of the dispersionless Toda lattice
hierarchy \cite{Dub,KO,AK}. Turning on every interaction term
$p^j\, (0\leq j\leq n-2)$ and using the technique of the pseudo differential
operator, it is shown that the $U(1)$ charge $q_\alpha$ spectrum of chiral
primary fields $\phi_\alpha$ becomes $\{q_\alpha =\alpha/(n-1)\, |\,
\alpha =0,1,\cdots,n-1\}$. This charge spectrum agrees with that conjectured
in \cite{CV}. In particular we have a marginal operator $\phi_{n-1}$ conjugate
to the coupling $t$ in (\ref{lax}). Furthermore, inspecting the $U(1)$
charge conservation we observe that the topological central charge is
$\hat c=1$ irrespective of $n$ and $c_1=n/(n-1)$ plays a similar role to ``the
first Chern class''. When $n=2$ we actually recover the $CP^1$ model.
This class of non scale-invariant LG models can be coupled to topological
gravity without any difficulty \cite{KO,AK}. However, we are still in the
regime $\hat c\ (=d)\leq 1$.

\vskip10mm

One of us (S.K.Y.) would like to thank T. Eguchi, K. Hori, H. Kanno and
T. Kawai for continual discussions on topological sigma models.
The work of S.K.Y. is supported in part by Grant-in-Aid
for Scientific Research on Priority Area 231 ``Infinite Analysis'',
the Ministry of Education, Science and Culture, Japan.

\newpage

%
%

\end{document}